# Testing the new QSM-6M optical module with the NEVOD Cherenkov water detector

**S.S. Khokhlov**[a,1]**, T.A. Karetnikova**[a]**, V.V. Kindin**[a]**, N.A. Pasyuk**[a]**, A.A. Petrukhin**[a] **and I.A. Shulzhenko**[a]

[a] *National Research Nuclear University MEPhI (Moscow Engineering Physics Institute)*
  *115409, Kashirskoe shosse 31, Moscow, Russia*
  *E-mail*: SSKhokhlov@mephi.ru

ABSTRACT: The method for studying characteristics of the response of optical modules of neutrino telescopes to various classes of events registered in the volume of the Cherenkov water detector NEVOD is discussed. Results of testing of an optical module with Hamamatsu R877 photomultiplier in single muon events and in events with high energy deposit are presented.

---

[1] Corresponding author.

## Contents

## 1. Introduction

In recent decades, Cherenkov water neutrino telescopes, such as IceCube [1], Baikal-GVD [2] and KM3NeT [3], are being actively developed for research in the field of neutrino physics and astrophysics. The main detecting elements of neutrino telescopes are optical modules. The capabilities of the telescope increase if its optical modules have an isotropic sensitivity in a $4\pi$ solid angle. Such module is called quasispherical and should consist of several photomultiplier tubes. The idea of a quasispherical module was first proposed in 1979 at the $16^{th}$ International Conference on Cosmic Rays (ICRC) [4]. Later, this idea was implemented when constructing the Cherenkov water detector (CWD) NEVOD [5], the basic element of which is a quasispherical module QSM-6 consisting of 6 PMTs.

Calibration of different optical modules under the same conditions is one of the important experimental problems of neutrino telescopes. Such calibration can be carried out at the Experimental complex NEVOD (Figure 1). The complex is based on the CWD with dimensions of $9\times9\times26$ m$^3$. On the bottom and on the top of the water tank, there are 80 scintillation counters of the calibration telescopes system, allowing the selection of near-vertical muons [6]. Around the CWD, the supermodules of the coordinate-tracking detector DECOR are installed. The DECOR detector allows precise reconstruction of near-horizontal muons [7].

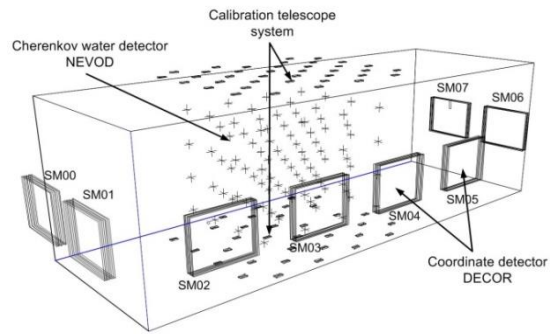

**Figure 1.** Structure of the Experimental complex NEVOD.

Independent detector systems for the muon track reconstruction provide opportunity to study the response of the optical module to such events [8]. In 2022, it is planned to calibrate the new optical module mDOM (multi-PMT Digital Optical Module) inside the volume of CWD NEVOD. The mDOM is being developed at the Muenster University (Germany) for the construction of IceCube-Upgrade detector and consists of 24 3-inch PMTs installed in a single housing [9].

To expand the detecting system of the CWD NEVOD, a new optical module QSM-6M is being developed. It consists of six Hamamatsu R877 PMTs directed along the axes of the orthogonal coordinate system. Hamamatsu R877 PMT has a flat photocathode with a diameter of 13 cm. The choice of this PMT is due to a high quantum efficiency and overall dimensions close to those of the currently used FEU-200 photomultipliers [10].

To test the technique for mDOM calibration, the optical module with Hamamatsu R877 photomultipliers was placed inside the CWD water volume.

## 2. Test results

From April 8 to July 8, 2021, data taking for studying QSM-6M response was carried out. During tests, the response to single near-vertical and near-horizontal muons was measured.

Figure 2 (left) shows the amplitude spectrum of module responses to muons with an average energy of 4 GeV, selected by a vertical telescope whose axis was parallel to the plane of the PMT photocathode and located at a 1-m distance from the optical module axis. The detection efficiency of the R877 PMT for telescope events is $90.2 \pm 0.2$ %. The mean response is $7.6 \pm 0.1$ photoelectrons (ph.e.).

The dependence of QSM-6M mean response on a distance between the module and muon track is shown in Figure 2 (right). Track events produced by near-horizontal muons with mean energy of 70-100 GeV were reconstructed with the DÉCOR detector. The module response was calculated as the sum of the responses of its PMTs. At distances large than 4 meters, the response curve begins to reveal features associated with the reflection of Cherenkov light from the water surface.

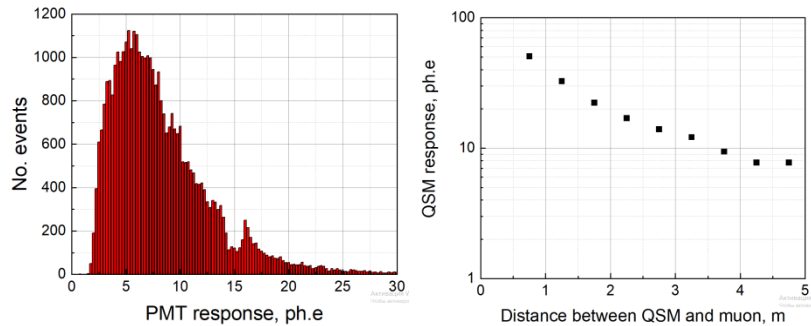

**Figure 2.** Response of PMT Hamamatsu R877 to near-vertical muons (left) and dependence of QSM-6M response on a distance to near-horizontal muons (right).

The CWD spatial lattice allows reconstructing the cascade curves of the showers generated by muons in the detector's sensitive volume [11]. The reconstruction technique is based on the amplitude responses of PMTs detecting the direct light from the cascade core (Figure 3, left). It is assumed that electrons and positrons produced in a shower move along the cascade axis and that the axis coincides with the muon track which can be measured in DECOR with the high precision. Figure 3 (right) shows an example of a reconstructed cascade curve of a shower with energy of 520 GeV: dots are the estimated numbers of charged particles calculated based on the responses of CWD PMTs; stars are the estimated numbers of charged particles calculated based on the responses of PMTs of QSM-6M (2 of 6 PMTs were used in the reconstruction); the curve is the Greizen approximation of cascade curve. As seen, results are in agreement with each other.

Based on the measured coordinates of muon tracks, we calculated the arrival direction of Cherenkov photons to the QSM-6M. Accounting for the responses of each of six PMTs of QSM-6M, the direction of Cherenkov light in individual event was reconstructed. Figure 4 (left) shows the distribution of events in the cosine of the angle between expected and reconstructed directions of Cherenkov light. The local minimum in the distribution is due to the presence of gaps between the DECOR supermodules. The mean value of the cosine is 0.88 with the standard deviation of 0.16.

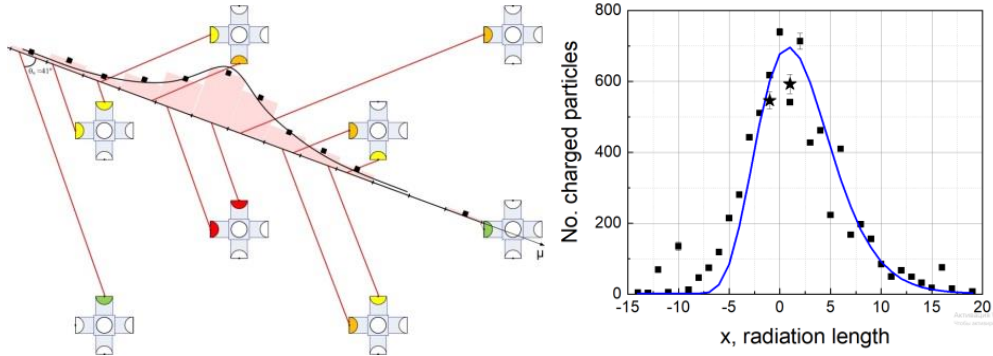

**Figure 3.** Reconstructed cascade curve of a shower detected in the CWD NEVOD.

Figure 4 (right) shows the dependence of the response of the QSM-6M on the arrival direction of the Cherenkov light. The dependence has been measured in events with distances between QSM-6M and near-horizontal muons in the range of 1-2 m. The minimal QSM-6M response is observed if the light arrives perpendicularly to the photocathode of any of the PMTs; the maximum of the response corresponds to the case when Cherenkov photons arrive under equal angles to three PMT at once. The non-sphericity of the QSM-6M, estimated as the ratio of the standard deviation and the mean response, equals to 14 %.

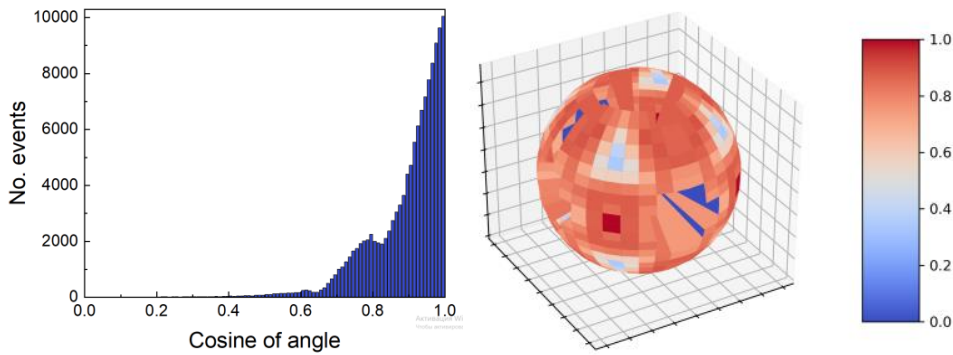

**Figure 4.** Distribution of events in the cosine of the angle between expected and reconstructed directions of Cherenkov light (left) and sphericity of QSM-6M response (right).

## 3. Conclusions

The Experimental complex NEVOD can be used as a test facility for calibration of optical modules of future neutrino telescopes. The technique of optical module calibration has been tested with the new module QSM-6M developed on the basis of the Hamamatsu R877 PMT. Calibration of the mDOM (optical module for IceCube-Upgrade) is planned for 2022.

## Acknowledgments

The work has been performed at the Unique Scientific Facility "Experimental complex NEVOD" with the support of the Ministry of Science and Higher Education of the Russian Federation, Project "Fundamental problems of cosmic rays and dark matter", No 0723-2020-0040.